\documentclass[twocolumn]{aastex62}

\usepackage{xspace}

\newcommand{\kzz}{\ensuremath{K_{zz}}}
\newcommand{\soo}{SO$_2$}
\newcommand{\coo}{CO$_2$}
\newcommand{\water}{H$_2$O}
\newcommand{\methane}{CH$_4$}

\newcommand{\update}{}

\shorttitle{Volatile-to-sulfur Ratios in WASP-39b}
\shortauthors{Crossfield}

\begin{document}

\title{Volatile-to-sulfur Ratios Can Recover a Gas Giant's Accretion History}

\correspondingauthor{Ian J.\ M.\ Crossfield}
\email{ianc@ku.edu}

\author{Ian J.\ M.\ Crossfield}
\affiliation{Department of Physics and Astronomy, University of
  Kansas, Lawrence, KS, USA}

\begin{abstract}
The newfound ability to detect \soo\ in exoplanet atmospheres presents
an opportunity to measure sulfur abundances and so directly test
between competing modes of planet formation.  In contrast to carbon
and oxygen, whose dominant molecules are frequently observed, sulfur
is much less volatile and resides almost exclusively in solid form in
protoplanetary disks. This dichotomy leads different models of planet
formation to predict different compositions of gas giant planets.
Whereas planetesimal-based models predict roughly stellar C/S and O/S
ratios, pebble accretion models more often predict superstellar
ratios.  To explore the detectability of \soo\ in transmission spectra
and its ability to diagnose planet formation, we present a grid of
atmospheric photochemical models and corresponding synthetic spectra
for WASP-39b (where \soo\ has been detected).  Our 3D grid
contains $11^3$ models (spanning 1--100$\times$ the solar abundance
ratio of C, O, and S) for thermal profiles corresponding to the
morning and evening terminators, as well as mean terminator
transmission spectra.
Our models show that for a WASP-39b-like O/H and C/H enhancement of
$\sim$10$\times$ Solar, \soo\ can only be seen for C/S and O/S
$\lesssim$\,1.5{\update $\times$ Solar}, and that WASP-39b's reported \soo\ abundance of
1--10~ppm may be more consistent with planetesimal accretion than with
pebble accretion models (although some pebble models also manage to
predict similarly low ratios). More extreme C/S and O/S ratios may be
detectable in higher-metallicity atmospheres, suggesting that smaller
and more metal-rich gas and ice giants may be particularly interesting
targets for testing planet formation models.  Future studies should
explore the dependence of \soo\ on a wider array of planetary and
stellar parameters, both for the prototypical \soo\ planet WASP-39b,
as well as for other hot Jupiters and smaller gas giants.

\vspace{1in}
\end{abstract}


\section{Introduction}
\label{sec:intro}

\subsection{Elemental Ratios and Planet Formation}

{\update Sulfur condenses into FeS at a condensation temperature of
$T_C$$\sim$660~K,} far higher than that of other volatiles commonly
observed in exoplanet atmospheres \citep[e.g.\ C, N, O, which all have
  $T_C$$\lesssim$180~K;][]{lodders:2003,wood:2019}.
Exoplanetary C,
N, and O abundances have therefore been frequently proposed as probes of whether
a given planet formed within or beyond the ``snow lines'' of various
C/N/O-bearing molecules \citep[e.g.,][]{oberg:2011,ohno:2023b}.

A longstanding example is a planet's carbon-to-oxygen ratio
\citep[C/O;][]{seager:2005}. These two elements, the most common in
the Sun after H and He \citep{lodders:2003,asplund:2009}, are expected
to form many of the dominant molecular species in gas giant
atmospheres.  CO, \water, \coo, and \methane\ can all induce prominent
spectral features and a planet's C/O should strongly affect the
relative abundances of these different molecules
\citep{seager:2005,madhusudhan:2012b,heng:2015c}. C/O was also the first
such ratio proposed to hold clues to a planet's formation and
evolution \citep[e.g.,][]{oberg:2011}, based on the idea that a gas
giant's composition should be determined by the location(s) in its
natal disk where the planet accretes most of its mass.

However,  growing evidence suggests that a planet's formation history
cannot be interpreted simply by reading off its C/O
ratio. For example, \cite{mordasini:2016} linked a chain of planet formation, disk,
and atmospheric models to find that a planet's C/O ratio may not
uniquely correlate with its initial formation location. Similarly,
subsequent studies of planet assembly also indicate that the C/O ratio
provides, at best, limited constraints on how and where a planet
formed and accreted most of its mass
\citep[e.g.,][]{turrini:2021,schneider:2021b,pacetti:2022,bitsch:2022}.

Other axes beyond C/O may therefore be necessary if we hope to
determine how and where a given planet may have formed. Although
numerous groups have explored the dependence of atmospheric nitrogen
abundance (as parameterized by N/O) on a planet's formation and
accretion, \citep{turrini:2021,schneider:2021b,ohno:2023a,ohno:2023b}
measuring a planet's N abundance is only feasible at temperatures
cooler than that of most hot Jupiters {\update (since at $T>$1000~K most N is
tied up in N$_2$)}.

\subsection{\soo\ in Gas Giant Atmospheres}

Sulfur's high $T_C$ implies that this species should be entirely in
the solid phase beyond $\sim$0.3~AU in protoplanetary disks
\citep[e.g.,][]{oka:2011}, where giant planet formation is thought to
occur.
At $T\gtrsim1000$~K, equilibrium chemistry predicts that most {\update atmospheric} sulfur
should reside in H$_2$S \citep{zahnle:2009,zahnle:2016,hobbs:2021,polman:2023,tsai:2023}.
However, the interaction of high-energy stellar photons with the
planet's atmosphere results in the H$_2$S abundance decreasing rapidly
at pressures $\lesssim 1$~mbar \citep{polman:2023,tsai:2023}.  Specifically,
 some H$_2$S is converted to \soo\  via photolysis of \water\ through  the
net reaction
\begin{equation}
\mathrm{H}_2\mathrm{S} + 2\ \mathrm{H}_2\mathrm{O} + \mathrm{photon} \rightarrow \mathrm{SO}_2 + 3\ \mathrm{H}_2 .
\end{equation}
This \soo\ resides at pressures of roughly 0.01--10~mbar, where it may
be observed via transmission spectroscopy if sufficiently abundant
\citep{polman:2023,tsai:2023}.  Because \soo\ contains three heavy
atoms, it may also be a useful probe of the average overall level of
metal enhancement in a planet's atmosphere \citep{polman:2023}.

Transmission spectroscopy of hot Jupiter WASP-39b through the JWST
Early Release Science program \citep[Program 1366;][]{batalha:2017ers}
revealed the clear signatures of numerous absorbers, including
\soo\ \citep{ers:2023,rustamkulov:2023,alderson:2023,ahrer:2023,feinstein:2023}.
Those observations detected excess absorption from 3.95--4.15\,$\mu m$
that was interpreted as roughly 1--10\,ppm of \soo\ at mbar pressures \citep{tsai:2023}.

In this paper, we explore how the abundances of S, as well as C and O,
determine the atmospheric \soo\ abundance and observable transmission
spectra of short-period, irradiated gas giants. Furthermore, we
suggest that \soo\ provides a unique opportunity to measure
volatile-to-sulfur ratios that could be a key discriminant between
competing planet formation theories.

We start by presenting a connection between planetary volatile-to-sulfur ratios
and planet formation models in Sec.~\ref{sec:formation}. In
Sec.~\ref{sec:models} we then present a new grid of photochemical
models and associated synthetic transmission spectra which we use to
investigate our ability to constrain atmospheric abundances via \soo. Finally, we conclude in Sec.~\ref{sec:conclusions}.

\section{Sulfur's Connection to Planet Formation}
\label{sec:formation}

As we describe below, the abundance ratios of volatiles to sulfur -- e.g.,
C/S and O/S -- may provide a powerful opportunity to distinguish between
competing models of planet formation.  Modern planet formation
simulations frequently track the atmospheric elemental abundances of
giant planets with a range of formation locations and migration
histories, and so provide hypotheses that can be tested via
measurements of atmospheric composition. However, these studies have
not yet focused specifically on sulfur.

Sulfur is thought to be carried largely in the volatile phase in the
ISM and during the earliest stages of protoplanetary disks, but is
quickly reprocessed until $\gtrsim$90\% of disk sulfur is carried in
solid (refractory) species \citep{kama:2019,legal:2021,riviere:2022}.
Its condensation temperature of $\sim$660~K \citep[][]{lodders:2003}
is high enough that sulfur remains in refractory form
throughout most of the disk, in contrast to volatile species such as C, N, and
O whose dominant carriers trace ``snow lines'' in the disk at
distances of several AU.

\begin{figure}
\centering
\includegraphics[width=0.48\textwidth]{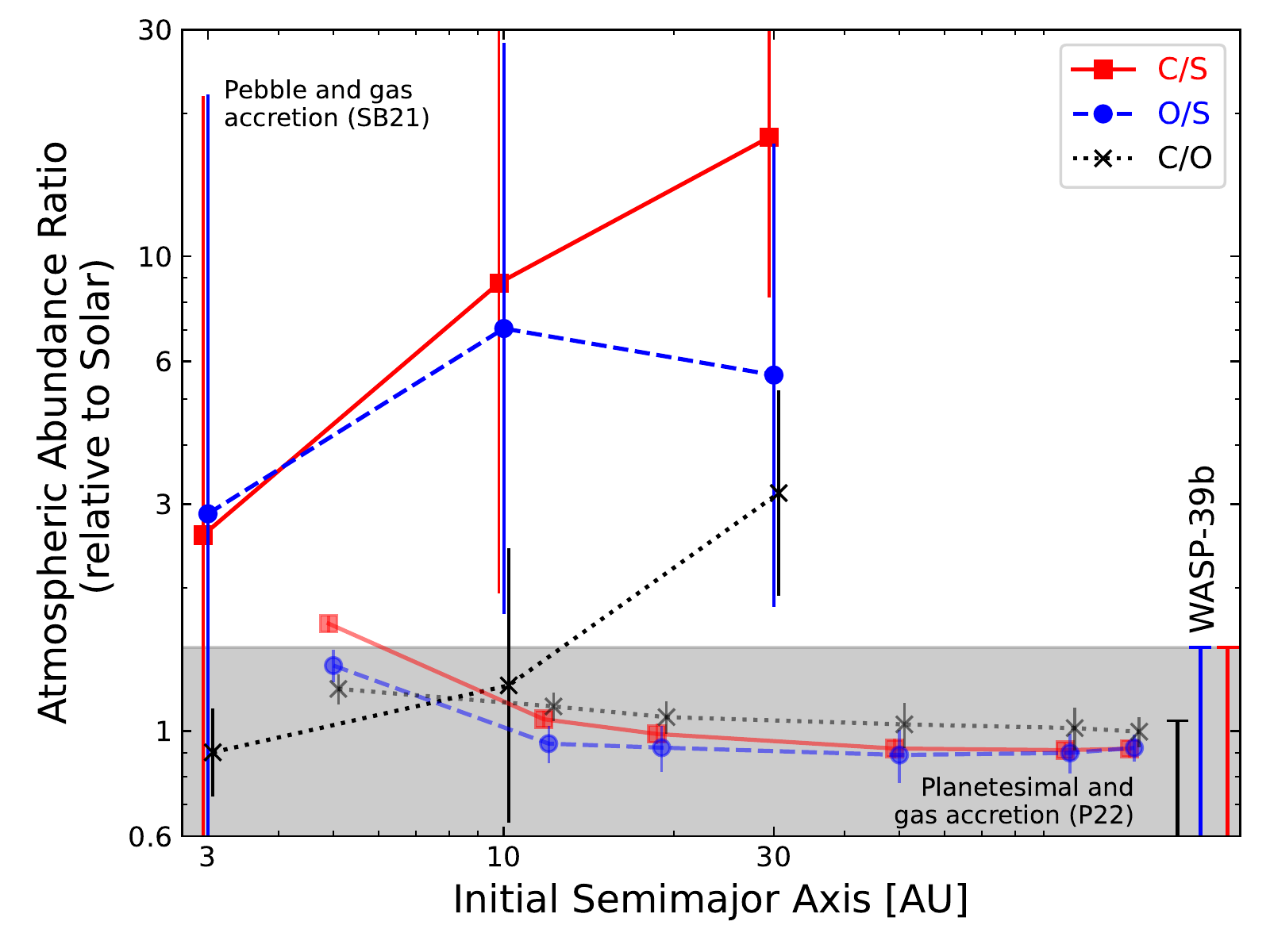}
\caption{Predicted atmospheric ratios of C/S, O/S, and C/O based on
  initial planet location, from the planet formation models of
  \citeauthor{schneider:2021b} (\citeyear{schneider:2021b}, SB21;
  upper, {\update thicker} curves) and \citeauthor{pacetti:2022}
  (\citeyear{pacetti:2022}, P22; lower, {\update thinner} curves). Although
  individual models span a range of final compositions (only the
  log-mean and standard deviation of each model set are plotted here),
  overall the differences suggest that volatile-to-sulfur ratios
  could provide a strong test as to whether a planet formed mainly via
  pebble or planetesimal accretion. The error bars at bottom-right and
  the shaded region show the abundance ratios inferred for WASP-39b
  from Fig.~\ref{fig:ratios}. }
\label{fig:formation}
\end{figure}

Pebble accretion models of planet formation predict that gas giants
may become highly enriched in volatile elements as compared to
refractories {\update \citep{schneider:2021a,schneider:2021b,schneider:2022}}. This occurs
because giant planets induce pressure extrema in the disk that inhibit
the inward migration and accretion of solids.  Thus in these models
less of the always-refractory S may be accreted when compared to
volatiles such as C or O (so long as the accreting planet is exterior
to the evaporation line of the dominant sulfur carrier{\update , FeS}).
Fig.~\ref{fig:formation} shows the predicted C/S, O/S, and C/O ratios
from the pebble accretion models of \cite{schneider:2021b}; for ease
of display, we present the mean and standard deviation (in log space)
of their predicted ratios over all combinations of viscosity parameter
$\alpha$ and refractory grain carbon content. Although their models
span a range of compositions, the general trend is that when pebble
accretion is the dominant mode for accreting solids, gas giants may
frequently exhibit roughly Solar C/O ratios but much higher C/S and
O/S ratios: from 3--20$\times$ Solar.

Models in which solids are accreted as planetesimals tell a different
tale.  Fig.~\ref{fig:formation} shows that the planetesimal-based
formation models of \cite{pacetti:2022} predict nearly Solar C/S, O/S,
and C/O ratios regardless of initial planet location (again, for
clarity we plot only the logarithmic mean and standard deviation of
their several models). These models \citep[consistent with and
  building on the initial models of][]{turrini:2021} predict higher
heavy-element abundances (C/H, etc.) for planets that started to form
closer in to the star, similar to the absolute abundance trend
predicted by pebble-accretion models \citep{schneider:2021b}. Thus the
absolute abundances of heavy elements in a planet's atmosphere may
also be a useful proxy for planet formation location.

Here we focus on the C/S and O/S ratios: Fig.~\ref{fig:formation}
shows that these ratios could allow a particularly powerful test of which
solid accretion mode dominated a planet's formation history.  Whereas
both types of models predict that a planet's final atmospheric C/O
ratio should be roughly stellar, pebble and planetesimal accretion
models can predict volatile-to-refractory (C/S and O/S) ratios that are
starkly different at all initial formation locations. In reality
planet formation may involve a combination of both planetesimal and
pebble accretion, in which case both these processes would shape the
observed composition of giant planets \citep{biazzo:2022}. Nonetheless
Fig.~\ref{fig:formation} still suggests that superstellar
volatile-to-sulfur ratios may be a compelling signpost of significant
pebble accretion.

Refractory elements less volatile
than sulfur have already been detected in some ultra-hot planets
\citep[e.g.,][]{lothringer:2021}, but these elements will condense in
most gas giant atmospheres.  Since sulfur only condenses at $\lesssim
660$~K \citep[at 0.1~mbar;][though in planetary atmospheres S vapor
  can exist at lower temperatures]{lodders:2003} its abundance -- and its
relation to that of volatile elements -- may thus be an especially
useful tool for constraining the dominant mechanism of planet
formation.


\section{Modeling}
\label{sec:models}

\subsection{Modeling Details}
Having shown that volatile-to-sulfur abundances may test planet formation theories, we
now explore how atmospheric S --- as measured by \soo\ abundances --- can reveal different
atmospheric compositions.
{\update Many factors have a strong impact on a planet's \soo\ abundance. E.g.,
\soo\ production is predicted to drop off steeply at $\lesssim$1000~K
\citep{polman:2023} and for FUV irradiation considerably higher than
that experienced by WASP-39b \citep{tsai:2023}; and aerosol formation could also deplete elements that would otherwise end up in photochemical \soo. Nonetheless, in this
study we restrict our} exploration to chemical composition and leave
these other axes to future studies.


We examine a three-dimensional parameter space of atmospheric
elemental abundances: $11^3$ combinations of a range of elemental
enhancements of carbon, oxygen, and sulfur (from solar to 100$\times$
solar), using the
\texttt{VULCAN}\footnote{\url{https://github.com/exoclime/VULCAN}}
photochemical kinetics code \citep{tsai:2017,tsai:2021}.  For each
combination of C, O, and S enhancement we calculate atmospheric
abundance profiles using \texttt{VULCAN}'s SNCHO chemical network,
which includes 575 chemical and photochemical reactions.  We use
WASP-39b as our simulated target; based on initial analyses of its
spectrum, we hold all abundances other than C, O, and S to 10$\times$
solar
\citep{ers:2023,rustamkulov:2023,alderson:2023,ahrer:2023,feinstein:2023,tsai:2023}.
The star WASP-39's XUV flux remains largely unknown, so we adopt the
same stellar spectrum used by \cite{tsai:2023}.  We compute our grid
using two different, pre-determined temperature profiles, one each for
the GCM-derived morning and evening terminator profiles presented by
\cite{tsai:2023}. {\update These thermal profiles are broadly
  consistent with the single profile recently retrieved by
  \cite{constantinou:2023}. Their retrieved, constant-with-altitude \soo\ abundances are broadly consistent with (if somewhat lower than) the detailed abundance profiles of \cite{tsai:2023} and of this work; further work is needed to compare abundances retrieved in this way with the predictions of photochemical codes}.  Finally, we also use the same
\kzz\ profile described by \cite{tsai:2023}, scaling with pressure as
$P^{-1/2}$ \citep[following][]{lindzen:1981,moses:2022}. The system
parameters and abundance values we used in our analysis are listed in
Table~\ref{tab:params}. All the \texttt{VULCAN} outputs are available
as machine-readable supplements to this
paper\footnote{\url{https://doi.org/10.5281/zenodo.7760360}}.

\begin{deluxetable*}{l l l l l}[bt]
\tabletypesize{\scriptsize}
\tablecaption{  Model Parameters: \label{tab:params}}
\tablewidth{0pt}
\tablehead{
\colhead{Name} & \colhead{Units} & \colhead{Description} & \colhead{Value} & \colhead{Source} 
}
\startdata
\multicolumn{5}{l}{\hspace{0.1in}\em System parameters:}\\
     $R_*$ &  $R_\odot$ & Stellar radius & 0.932  &  Carter \& May, et al., in prep.\\
     $R_P$ &  $R_J$ & Planetary radius & 1.279 &  Carter \& May, et al., in prep.\\
     $M_P$ &  $M_J$ & Planetary mass & 0.281 &  Carter \& May, et al., in prep.\\
     $g_P$ &  m~s$^{-2}$ & Planetary surface gravity & 4.26  &  Carter \& May, et al., in prep.\\
$a$ &    AU &  Semimajor axis & 0.04828 & Carter \& May, et al., in prep.\\
\multicolumn{5}{l}{\hspace{0.1in}\em Modeling parameters:}\\
He/H  & -- & Solar volume mixing ratio & $8.38 \times 10^{-2}$  & \cite{lodders:2020}\\
C/H  & -- & Solar volume mixing ratio & $2.95 \times 10^{-4}$  & \cite{lodders:2020}\\
O/H  & -- & Solar volume mixing ratio & $5.37 \times 10^{-4}$   & \cite{lodders:2020}\\
S/H  & -- & Solar volume mixing ratio & $1.41 \times 10^{-5}$   & \cite{lodders:2020}\\
$P$ & bar & Pressure range & \multicolumn{2}{l}{10--10$^{-9}$  } \\
$P_0$ & bar & Reference pressure & \multicolumn{2}{l}{0.01  } \\
$z$ & deg & Zenith angle & \multicolumn{2}{l}{83} \\
C  & -- & Abundance relative to Solar & \multicolumn{2}{l}{1, 1.8, 3, 5.6, 7.5, 10, 13, 18, 30, 56, 100} \\
O  & -- & Abundance relative to Solar & \multicolumn{2}{l}{1, 1.8, 3, 5.6, 7.5, 10, 13, 18, 30, 56, 100} \\
S  & -- & Abundance relative to Solar & \multicolumn{2}{l}{1, 1.8, 3, 5.6, 7.5, 10, 13, 18, 30, 56, 100} \\
\enddata
\end{deluxetable*}

We then use the
\texttt{petitRadTrans}\footnote{\url{https://petitradtrans.readthedocs.io/}}
radiative transfer code \citep{molliere:2019} to calculate synthetic
transmission spectra corresponding to each \texttt{VULCAN} run. We
convert \texttt{VULCAN}'s volume mixing ratios (VMRs) to
\texttt{petitRadTrans}' mass mixing ratios.  We use
\texttt{petitRadTrans}' medium-resolution, correlated-k opacity
sources, giving our synthetic spectra a {\update resolving power} of
$\sim$1,000 from 1--25\,$\mu m$. The molecules and opacity sources we
include are \water, CO, \coo, \soo, \methane, HCN, H$_2$S, CH$_3$,
C$_2$H$_2$, C$_2$H$_4$, CN, CH, OH, and SH
\citep{rothman:2010,polyansky:2018,yurchenko:2020,underwood:2016,yurchenko:2014,harris:2006,barber:2014,chubb:2018,azzam:2016,adam:2019,chubb:2020,brooke:2014,bernath:2020,syme:2020,masseron:2014,brooke:2016,yousefi:2018,gorman:2019}.
We have three sets of synthetic spectra: one each for the morning and
evening thermal profiles, plus a set of spectra that are the
mean of the morning and evening spectra (corresponding to the
transmission spectrum that would be observed during transit). All the
\texttt{petitRadTrans} synthetic spectra are also available as
machine-readable supplements to this paper\footnote{\url{https://doi.org/10.5281/zenodo.7760360}}.

We note that the abundances of 15 elements (including C and O, but not
S) were measured for WASP-39 \citep{polanski:2022}, revealing a
composition for which every element is consistent with the solar
values at $<1\sigma$.  Similarly, the C/O ratio of $0.46 \pm 0.09$ is
consistent with the solar value of 0.55 \citep[assuming the abundances
  of][]{lodders:2020}.  Therefore, although generally stellar (not
solar) abundances are the appropriate referent for atmospheric
modeling, we elect to use abundance levels scaled from Solar values
given WASP-39's chemical similarity to the Sun {\update \citep{polanski:2022}.}

{\update Finally, we also note that for a nominal WASP-39b model with an
atmosphere enhanced to 10$\times$ Solar abundances, the transmission
contribution function in regions of maximal \soo\ absorption span a
pressure range of roughly 0.01--30~mbar. Although a weighted average of
the \soo\ concentration (e.g., via weighting by atmospheric mass
density) may give a somewhat more representative view of the peak of
the photochemical \soo\ abundance, for simplicity \citep[and to facilitate
direct comparison with the work of][]{tsai:2023} here we report
the average of our modeled \soo\ concentration from 0.01--10~mbar.}

\begin{figure}
\centering
\includegraphics[width=0.45\textwidth]{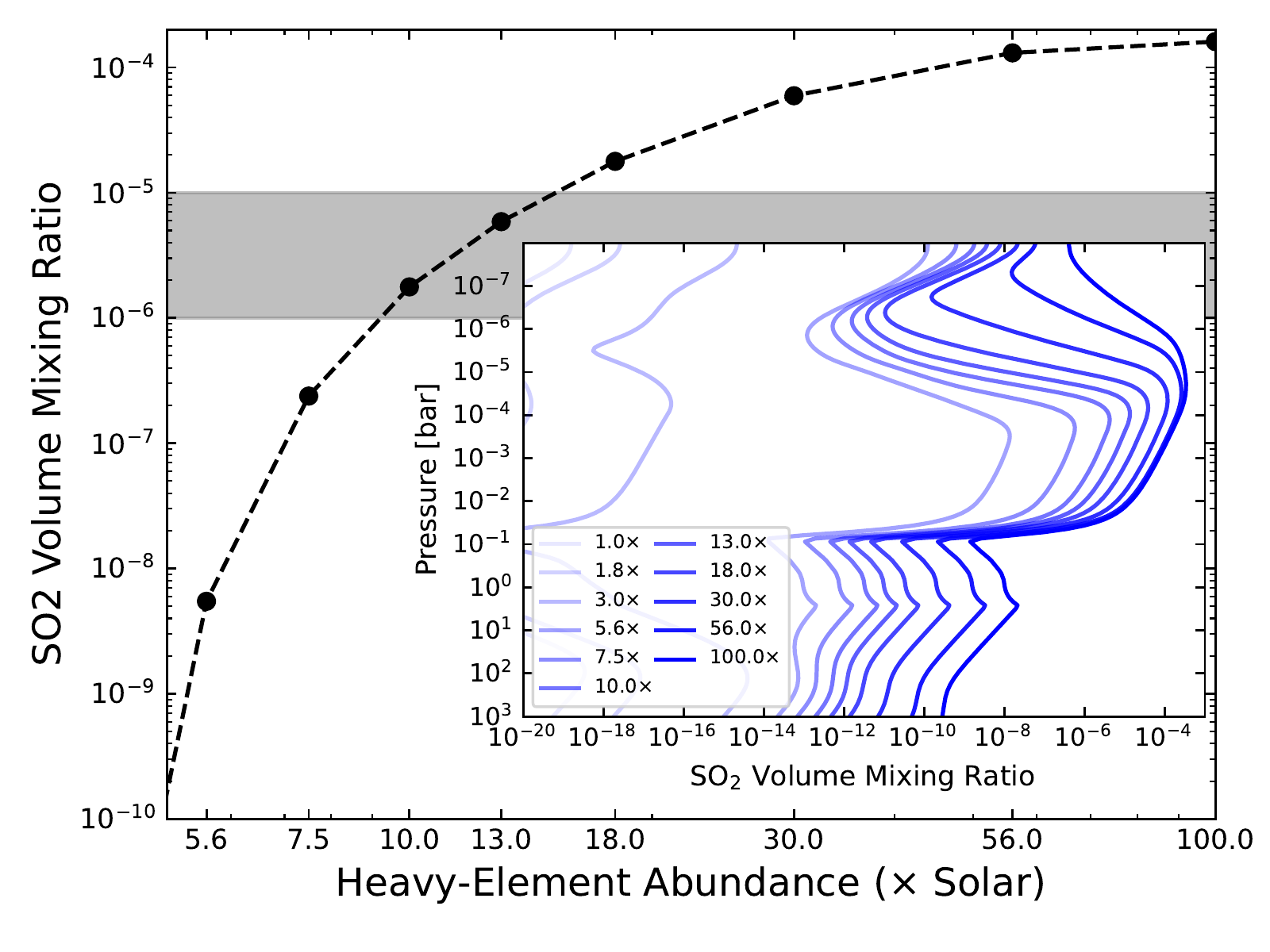}
\caption{In the main panel, the dashed line shows the \soo\ VMR,
  averaged from 0.01--10~mbar, as all elemental abundances are
  increased in lockstep; the gray region shows the approximate
  \soo\ abundance reported for WASP-39b. The inset shows the
  full vertical \soo\ profiles. }
\label{fig:metallicity}
\end{figure}

\begin{figure}
\centering
\includegraphics[width=0.48\textwidth]{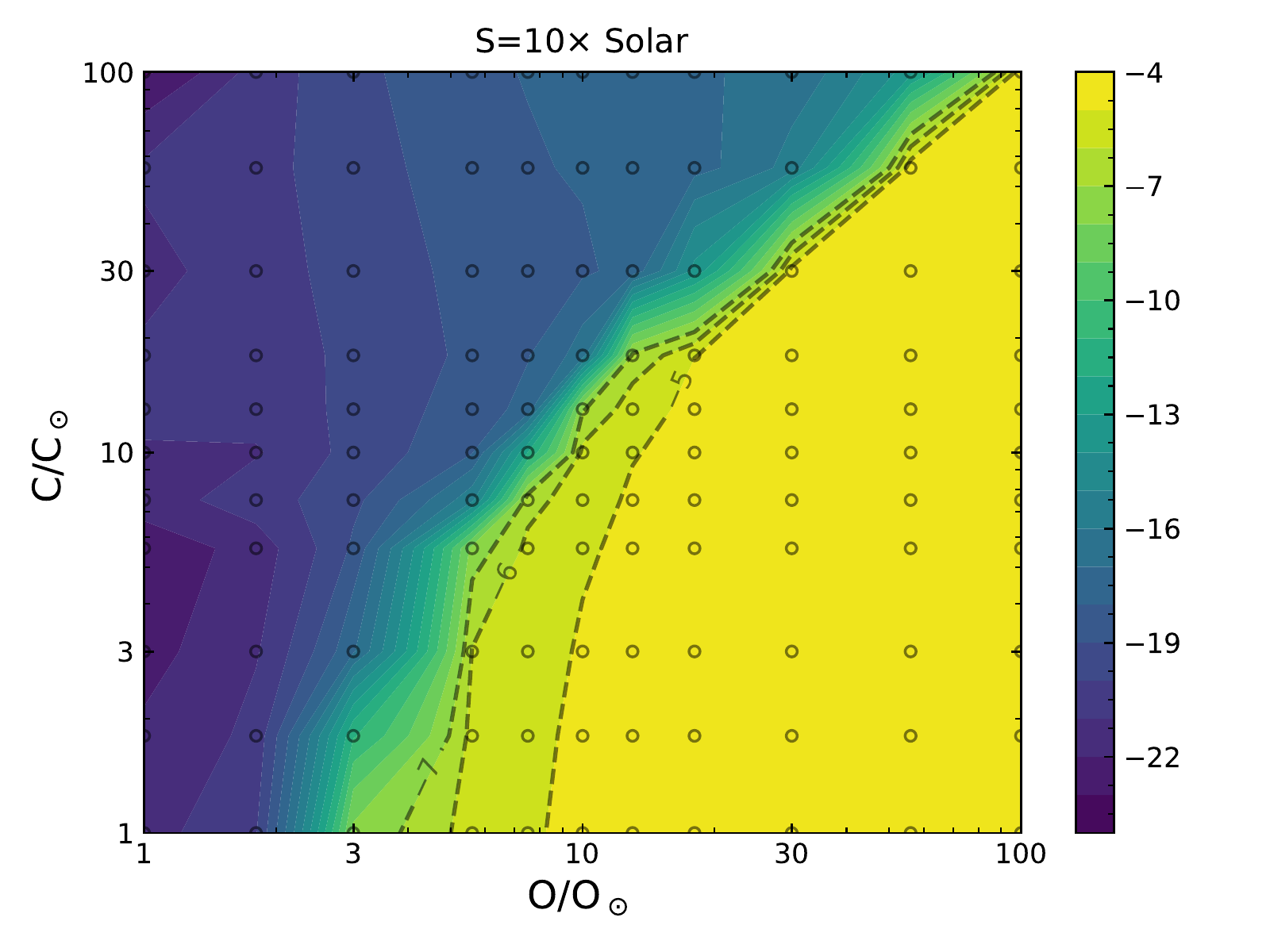}
\includegraphics[width=0.48\textwidth]{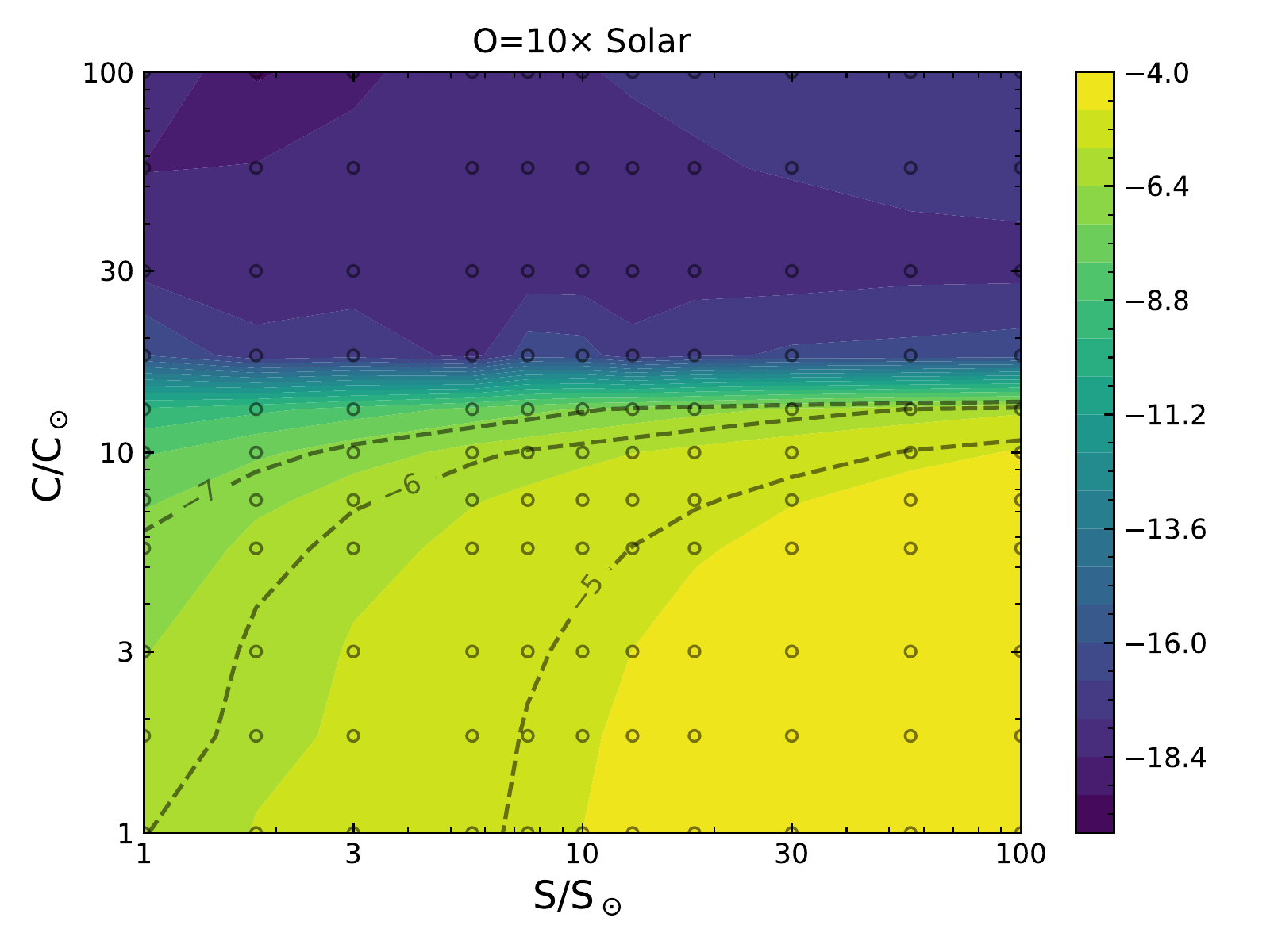}
\includegraphics[width=0.48\textwidth]{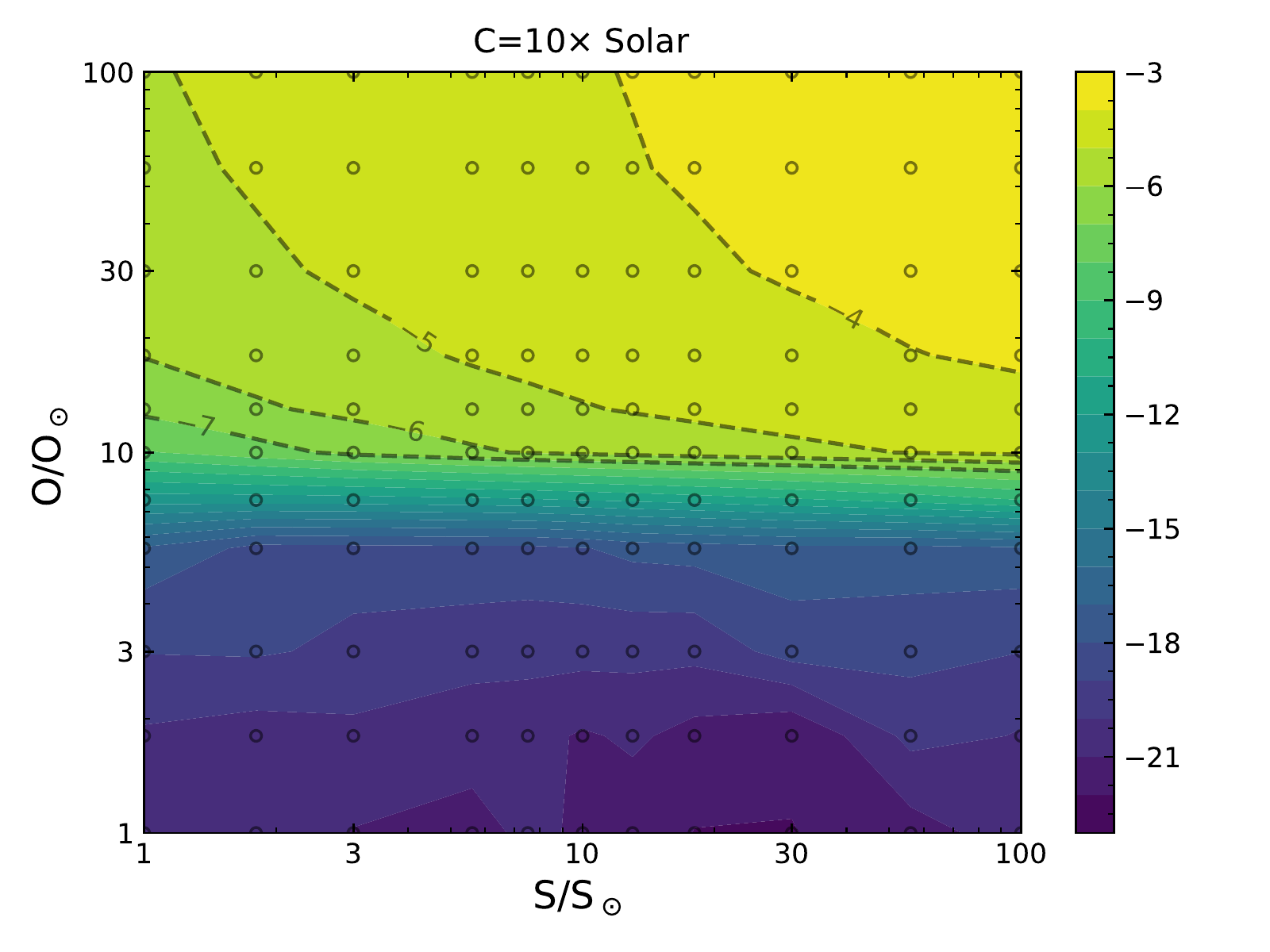}
\caption{Abundance of \soo\ (averaged from 0.01--10~mbar) as
  combinations of two elemental abundances are varied: C vs.\ O ({\em
    top}), C vs.\ S ({\em middle}), and O vs.\ S ({\em bottom}).  The
  points indicate the locations of the model grid points, which we
  linearly interpolate between. In each panel, all other elements are
  held fixed at 10$\times$ Solar abundance. }
\label{fig:grids}
\end{figure}

\begin{figure}
\centering
\includegraphics[width=0.48\textwidth]{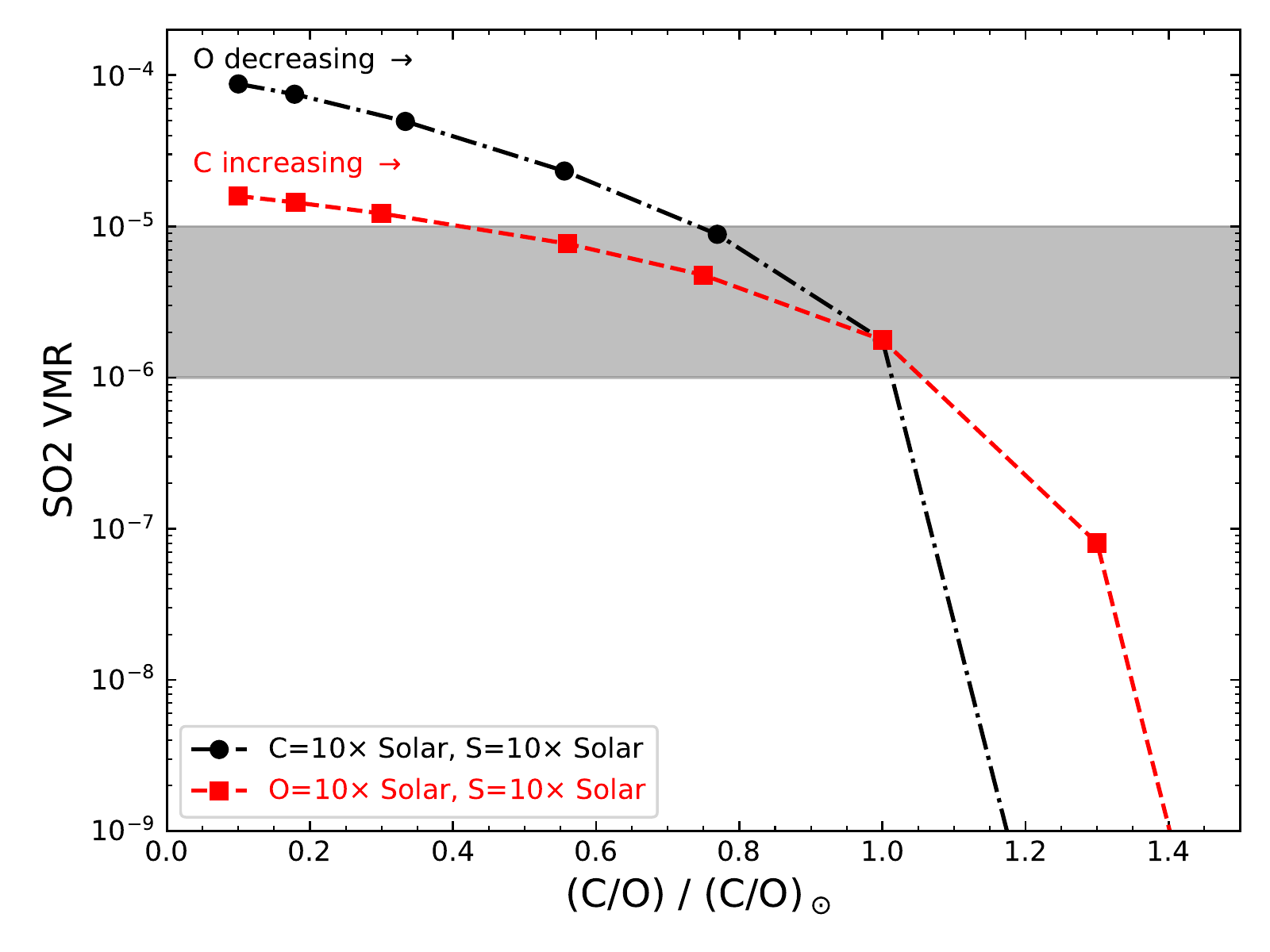}
\includegraphics[width=0.48\textwidth]{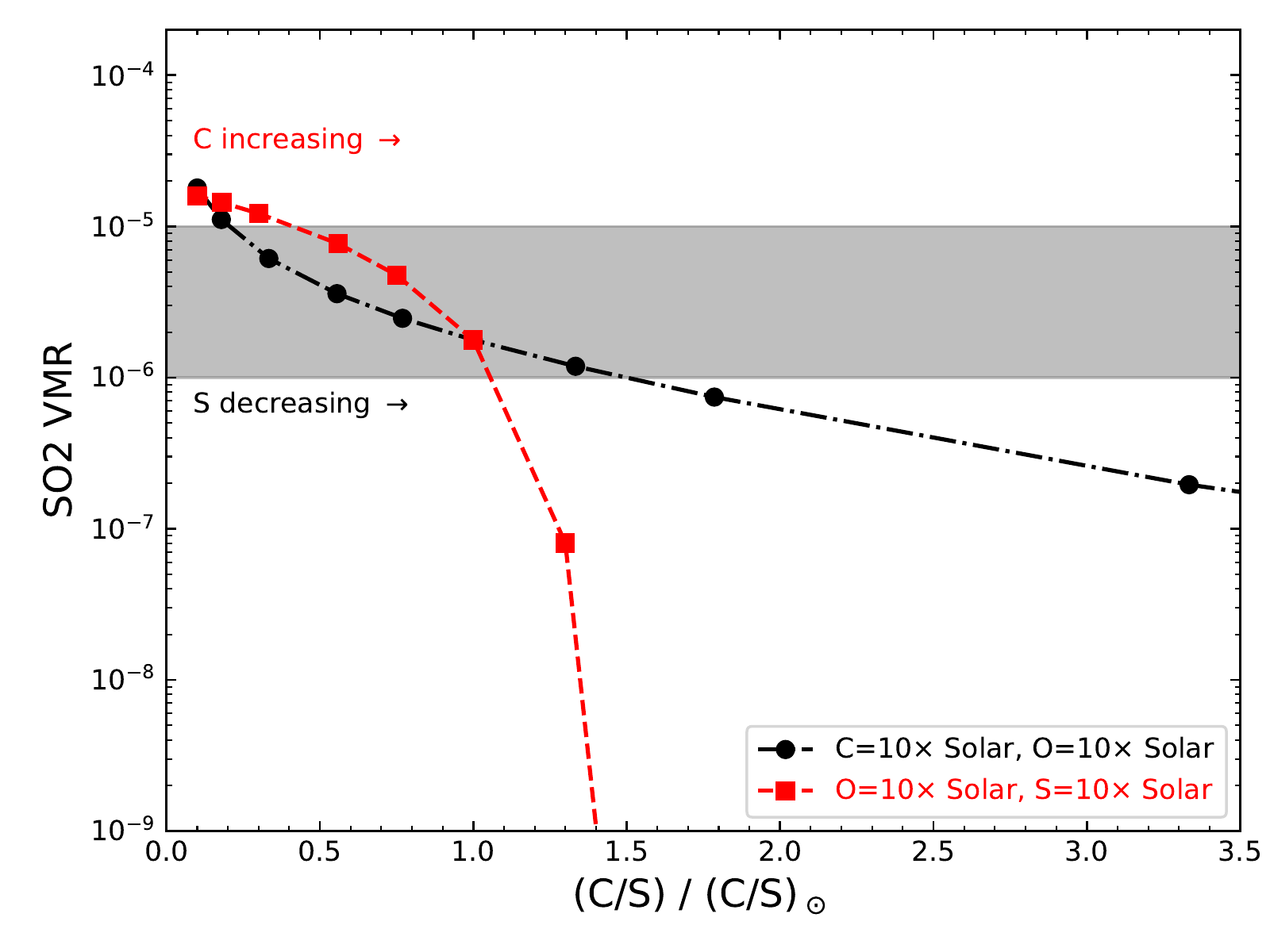}
\includegraphics[width=0.48\textwidth]{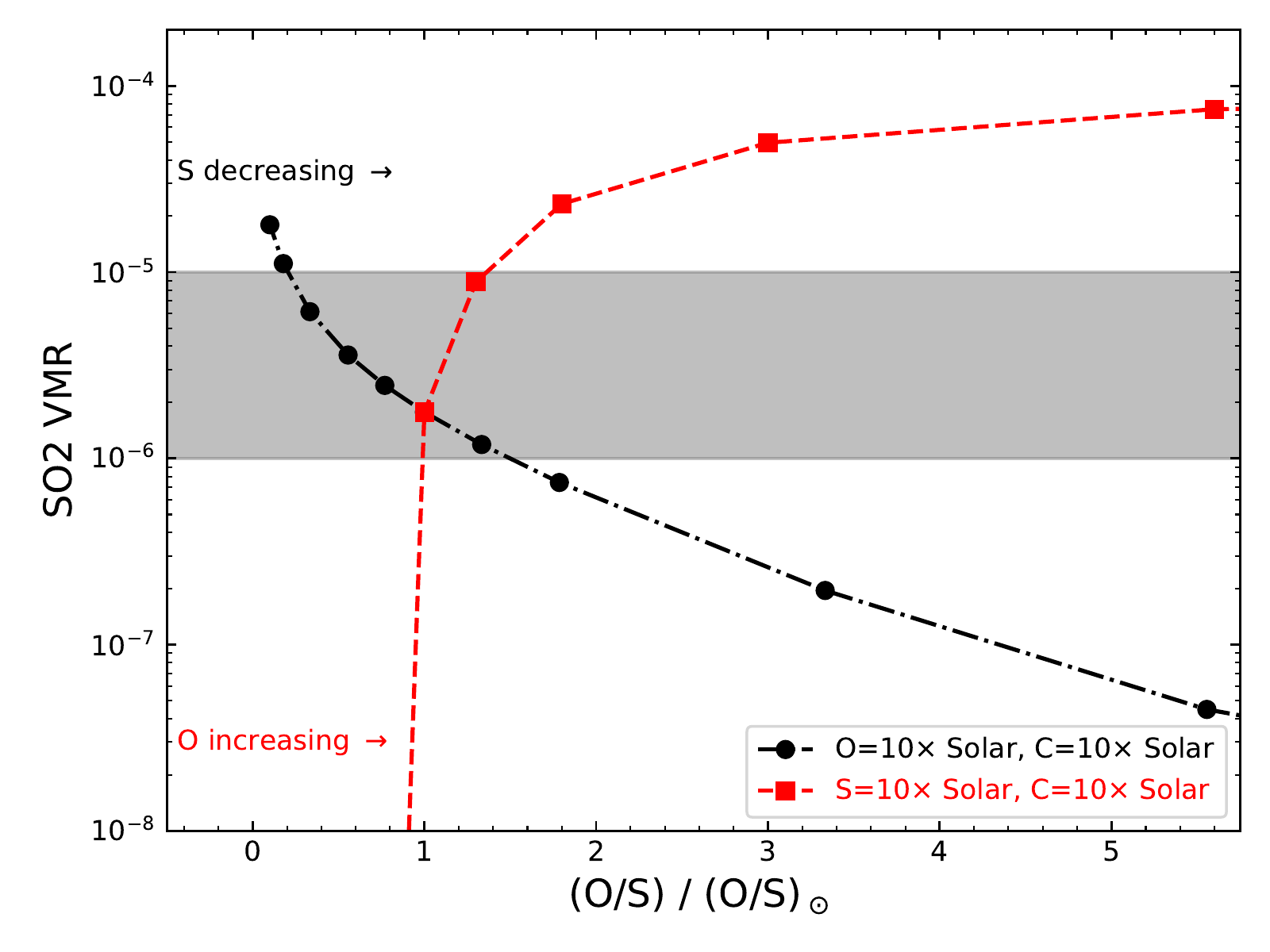}
\caption{Abundance of \soo\ (averaged from 0.01--10~mbar) as the
  atmospheric ratios of C/O ({\em top}), C/S ({\em middle}), and O/S
  ({\em bottom}) are varied.  The gray regions show the approximate
  \soo\ abundance reported for WASP-39b, which correspond here to
  C/S and O/S $\lesssim$1.5. }
\label{fig:ratios}
\end{figure}

\begin{figure}
\centering
\includegraphics[width=0.48\textwidth]{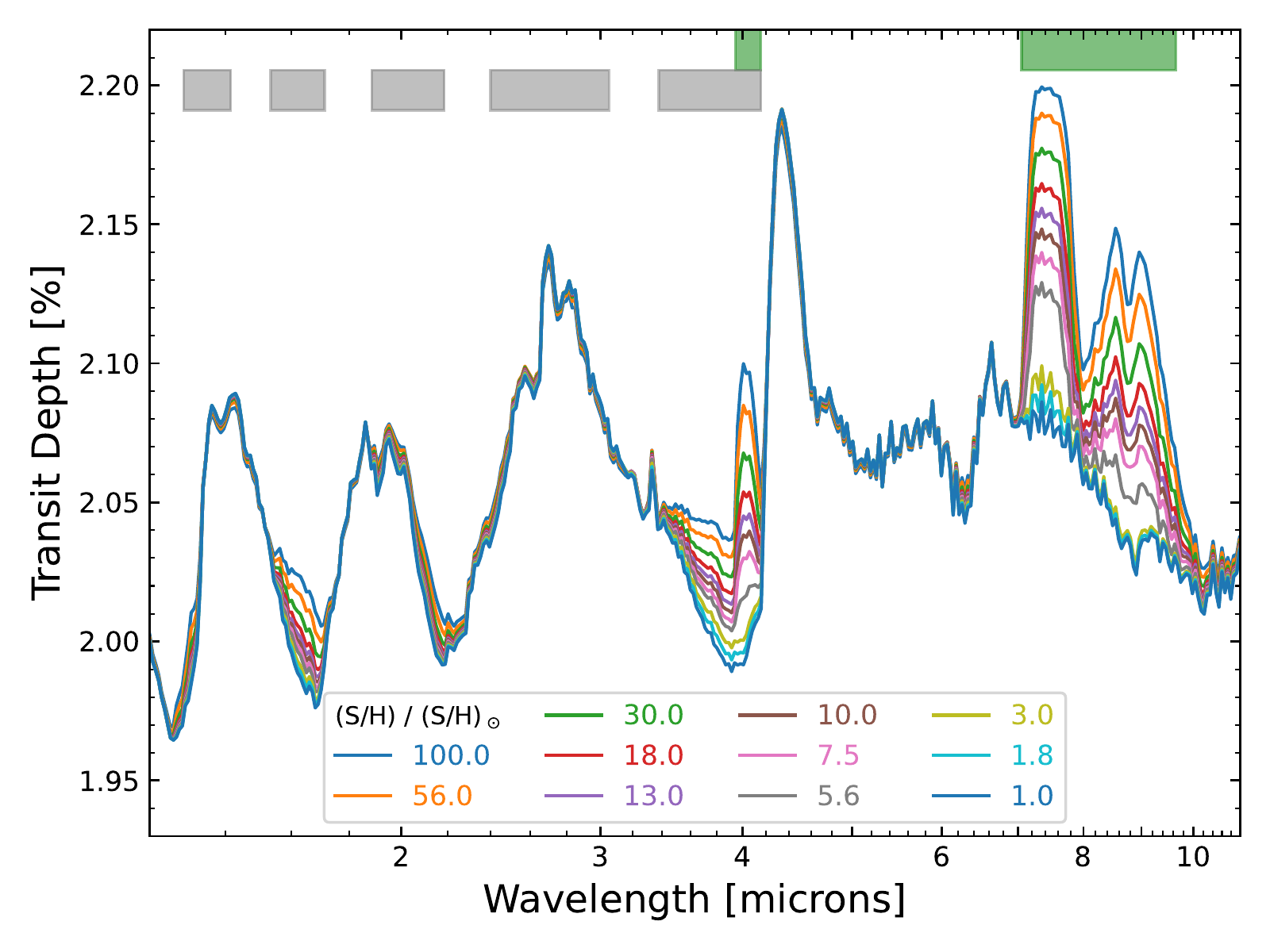}
\caption{Synthetic spectra showing the effect of varying the S/H ratio
  from 1--100$\times$ Solar while keeping O/H and C/H at 10$\times$
  Solar, as inferred for WASP-39b (equivalent to varying C/S and O/S
  from 10--0.1$\times$ Solar). At 10$\times$ Solar metallicity
  \soo\ is detectable up to volatile-to-sulfur ratios of
  $\lesssim$3$\times$ Solar. The shaded rectangles show where
  absorption is dominated by \soo\ (green) and by H$_2$S (grey). }
\label{fig:spectra1}
\end{figure}

\subsection{Differences From Previous Studies}
Our modeling effort expands on previous exploration of 
\soo\ in several ways.  The first such study presented a
comprehensive examination of \soo\ abundance in hot Jupiter
atmospheres \citep{polman:2023}. Their study also used the
\texttt{VULCAN} code to cover metallicities of 1--20$\times$ Solar,
three values of \kzz\ (constant with altitude, but spanning three
orders of magnitude), C/O ratios from 0.25--0.9, several different
stellar spectra, and three planetary temperatures (spanning 400~K).
More recently, \cite{tsai:2023} demonstrated that \soo\ causes the
4.2\,\micron\ absorption feature seen in JWST spectroscopy of WASP-39b
\citep{rustamkulov:2023,ahrer:2023,alderson:2023}. That investigation
used four photochemistry codes (including \texttt{VULCAN}) to span
three metallicities (from 5 to 20$\times$ Solar), three \kzz\ profiles
(spanning two orders of magnitude), three C/O ratios (from 0.25 to
0.75), three stellar spectra (spanning two orders of magnitude in
irradiation), and a range of temperatures (from 600 to 2000~K).

Our analysis builds upon both these works by (i) examining a more
densely sampled and fully two-dimensional grid of C and O abundances,
(ii) adding a third dimension by exploring a wide range of S
abundances, and (iii) extending the analysis up to significantly
higher atmospheric metallicities.

\subsection{Discussion and Interpretation}

Fig.~\ref{fig:metallicity} shows how the \soo\ abundance increases as
all three elements (C, O, and S) are increased in lockstep. Whereas
the abundance of the triatomic \coo\ (produced via equilibrium
chemical processes) increases quadratically with metallicity
\citep{zahnle:2009a}, the more complicated formation pathways of
\soo\ result in a more complicated metallicity dependence: much
steeper than \coo\ at low metallicity, and shallower at high
metallicity. This suggests that if high-metallicity ($>100\times$
Solar) ice giants also form photochemical \soo, its VMR is 
unlikely to be $\gtrsim$200~ppm \citep[as found by][]{tsai:2021}.

Fig.~\ref{fig:grids} shows three slices through our 3D abundance grid,
depicting the \soo\ volume mixing ratio (VMR, averaged from
0.01--10~mbar) versus C, O, and S abundances.  Fig.~\ref{fig:ratios}
shows 1D slices as plotted against the abundance ratios C/O, C/S, and
O/S.

Fig.~\ref{fig:ratios} reveals that the \soo\ abundances decreases as the
atmospheric C/O ratio increases (while the S abundance is held
constant). Consistent with the results of \cite{polman:2023}, the
dependence on C/O is similar regardless of whether we increase C or
decrease O (though with a slightly steeper dependence when O is
varied). In either case, when C/O increases beyond the Solar value the
\soo\ abundance rapidly drops below detectable levels.  The detection
of \soo\ is therefore a strong sign of a C/O ratio $\lesssim$ the Solar value.

\begin{figure}
\centering
\includegraphics[width=0.48\textwidth]{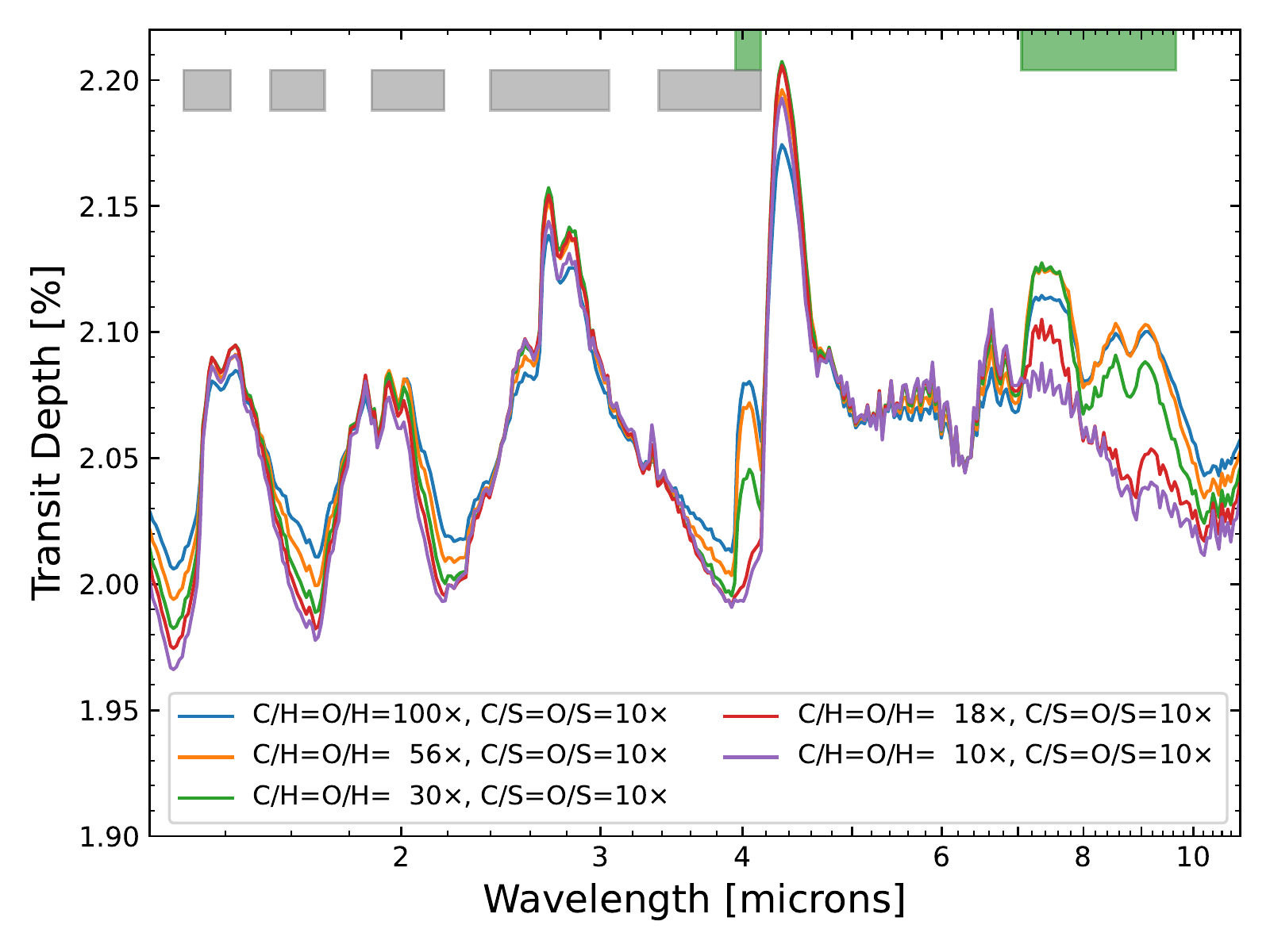}
\caption{Synthetic spectra showing the effect of varying C/H and O/H
  from 10--100$\times$ Solar while keeping the C/S and O/S ratios at
  10$\times$ Solar, as might be expected from pebble accretion
  (equivalent to varying S/H from 1--10$\times$ Solar). At 10$\times$
  Solar C/S and O/S, \soo\ is detectable down to a volatile
  enhancement level of $\sim$18$\times$ Solar. The shaded rectangles
  show where absorption is dominated by \soo\ (green) and by H$_2$S
  (grey). Note that for volatile enrichment $\gtrsim$100$\times$ Solar
  the amplitude of spectral features begins to decrease as the mean
  molecular weight begins to increase.}
\label{fig:spectra2}
\end{figure}

More excitingly, Fig.~\ref{fig:ratios} shows how measurements of the
\soo\ abundance can distinguish between a variety of
volatile-to-sulfur ratios. JWST transit spectroscopy of WASP-39b
reveal it to have an atmospheric metallicity $\sim$10$\times$ Solar, a
C/O ratio $\lesssim$ the Solar value, and a \soo\ VMR of 1--10~ppm
\citep{ers:2023,rustamkulov:2023,alderson:2023,ahrer:2023,feinstein:2023,tsai:2023}. Assuming an overall metallicity of 10$\times$ Solar, the bottom two panels of
Fig.~\ref{fig:ratios} show that the \soo\ measurement constrains both
C/S and O/S to $\lesssim$1.5$\times$ Solar. Reference to
Fig.~\ref{fig:formation} demonstrates that such values are rather more
consistent with planetesimal accretion models \citep{turrini:2021,pacetti:2022}
than with pebble accretion \citep{schneider:2021b}.  The only pebble formation models of \cite{schneider:2021b} that can approximately reproduce these ratios assume an initial formation location of 3~AU and $\alpha \le 5 \times 10^{-4}$.

We also show a few representative examples of our synthetic
transmission spectra.  In Fig.~\ref{fig:spectra1} we see the effect on
WASP-39b's transmission spectra of varying the S/H ratio from
1--100$\times$ Solar, while keeping O/H=C/H=10$\times$ Solar (as
inferred for WASP-39b); this is equivalent to varying C/S and O/S from
10$\times$ down to 0.1$\times$ Solar.  The figure shows that at
WASP-39b-like volatile enrichment levels, \soo\ has a significant
impact on a planet's transmission spectrum for S/H$\gtrsim3\times$
Solar (C/S or O/S $\lesssim$3$\times$ Solar).

Lest one {\update conclude from Figs.~\ref{fig:ratios} and~\ref{fig:spectra1}
 that high volatile-to-sulfur ratios $\gtrsim$3$\times$ Solar are
impossible to detect}, Fig.~\ref{fig:spectra2} shows that \soo\ is easily
discernible at C/S=O/S=10$\times$ Solar --- so long as the atmospheric
volatile enrichment {\update (i.e., O/H and C/H) is} $\gtrsim$18$\times$ Solar.  Since smaller,
lower-mass planets are thought to form with more metal-rich
atmospheres \citep[e.g.,][]{fortney:2013}, such planets may therefore
be the best targets to look for \soo\ to reveal the extreme
volatile-to-sulfur ratios that would be a signpost of formation via
pebble accretion.


\section{Conclusions }
\label{sec:conclusions}

\subsection{Summary}
In conclusion, we have shown that a planet's formation and migration
history may imprint unique elemental abundance signatures on the
planet's atmosphere: specifically, on the C/S and O/S ratios.  In
particular, Fig.~\ref{fig:formation} and Sec.~\ref{sec:formation}
demonstrate how formation models in which solids are predominately
accreted as pebbles often predict much higher volatile-to-sulfur ratios (C/S
and O/S) than models in which  solids are accreted as planetesimals. Although a
planet's N/S ratio has also been proposed as a diagnostic of formation
\citep{turrini:2021,pacetti:2022}, detecting both these species in a
gas giant may be challenging: S is mostly only visible as \soo\ at
$T\gtrsim1000$~K \citep{polman:2023} while N is mostly detectable as
NH$_3$ at $T\lesssim1000$~K \citep{ohno:2023b}. Thus
the C/S and O/S  ratios may be more generally useful (and measurable) than N/S.

We then present a grid of photochemical models and synthetic spectra
for WASP-39b in which C, O, and S abundances are varied to determine
the impact on the observable \soo\ signature (Sec.~\ref{sec:models}
and Fig.~\ref{fig:grids}).  {\update Consistent with previous studies, we find
that \soo\ should be most prevalent at high metallicity; we also find
that it should be more abundance at higher elemental O and S
abundances but lower C abundances.}
By examining how the \soo\ abundance
varies with C/S and O/S in our models, Fig.~\ref{fig:ratios} shows that
WASP-39b's atmospheric \soo\ abundance may constrain C/S and O/S to
levels more consistent with planetesimal accretion than pebble
accretion -- although some pebble accretion models also predict lower volatile-to-sulfur ratios that seem more consistent with WASP-39b.

Additional work is certainly needed to confirm this conclusion. The
main planet formation models discussed in Sec.~\ref{sec:formation} all
result in planets more massive than WASP-39b, which suggests the need
for caution in linking these models to WASP-39b's composition.  Also,
relatively few planet formation studies report the final planetary C,
O, and S abundances --- additional, independent models tracking these
abundances and featuring either planetesimal or pebble accretion would
be extremely useful to understand the robustness of the dichotomy in
volatile-to-sulfur ratios shown in Fig.~\ref{fig:formation}.

\subsection{Future Work}
There is also considerable scope for further exploration of the
planetary and stellar parameters that impact the observed
\soo\ abundance.  These include bolometric irradiation, received XUV
flux, aerosols, additional T-P profiles, better high-temperature
opacity data in the UV, and the observability of \soo\ or other
S-bearing species via thermal emission --- not to mention the impact
of additional species such as nitrogen.

Regarding high-energy stellar flux, \cite{hobbs:2021} reported that
increasing UV flux by up to two orders of magnitude merely results in
the \soo\ abundance profile being pushed to somewhat deeper pressures,
but with the overall abundance relatively unchanged.  On the other
hand, \cite{tsai:2023} instead find that increasing the NUV flux
minimally changes the \soo\ profile, while increasing the FUV flux can
decrease the \soo\ abundance by $\sim10^4$. If the influence of UV
flux could be pinned down, it would provide strong motivation for
measuring the intrinsic spectra of stars with planets expected to
exhibit \soo\ (or other) photochemistry.

Also regarding WASP-39's intrinsic stellar spectrum, we note that [S/H] has
still not been measured for this star. Although the star's generally
Solar-like elemental abundances \citep{polanski:2022} suggest that its
sulfur abundance may also be nearly Solar, [S/H] should be
measured for WASP-39 as it has been for many other planet-hosting stars \citep[e.g.,][]{costasilva:2020} to confirm the expected, Solar-like sulfur abundance.

In this work we have not explored planetary nitrogen abundance, but it
may also be an important axis.  For example, the nitrogen-to-oxygen
ratio in gas and solids should vary by several orders of magnitude
from 1--100~AU \citep{ohno:2023a}. And if planet formation is
dominated by planetesimal accretion, then the sulfur-to-nitrogen ratio
may be a useful probe of whether a gas giant accreted most of its mass
via solids or gas \citep{turrini:2021}.

Further exploration is also warranted into the effect of different
\kzz\ values on the observable \soo\ abundance.  \cite{tsai:2023}
found that varying their \kzz\ profile (nominally spanning $5 \times
10^7$ to $10^{11}\mathrm{~cm}^2\mathrm{\ s}^{-1}$) by $\pm$ an order
of magnitude had only a minor impact, consistent with the results of
\cite{hobbs:2021} from isothermal models with \kzz\ spanning
$10^6$--$10^{12}\mathrm{~cm}^2\mathrm{\ s}^{-1}$.  In contrast,
other studies  report that increasing a constant-with-altitude
\kzz\ to $10^{11}\mathrm{~cm}^2\mathrm{\ s}^{-1}$ sharply decreases
the amount of observable \soo\ \citep{tsai:2021,polman:2023}. Although \kzz\ is a challenging
quantity to empirically constrain, it would at least be useful to
understand how it quantitatively impacts the measurement of \soo\ in
planetary atmospheres.

{\update Future} studies might also test the impact of
self-consistently modeling the thermal and chemical profiles in the
atmospheres, thereby accounting for thermal back-reaction of
photochemical species such as \soo\ on the planet's vertical
temperature structure.  Similarly, the interplay of global circulation
and atmospheric chemistry may reveal that predictions made from 1D
models (as in this work) -- or even from post-processed chemistry-free
global circulation models -- may lead to inaccurate interpretations of
exoplanet measurements \citep{lee:2023}.

{\update JWST/NIRSpec may be the ideal instrument to best constrain C, O, and S
and so begin to test the connection between atmospheric \soo\ and
planet formation posited in this paper. Even the G395 modes (let alone
the Prism!) cover multiple strong bands of \coo, CO, CH$_4$, and
\water, as well as the \soo\ described by \cite{tsai:2023}.  The
signal of \soo\ may be even stronger in the MIRI/LRS bandpass, but the
overall S/N may be lower there -- in addition, shorter wavelength
coverage may still be needed to provide a complete elemental assay;
the feasibility of MIRI for \soo\ measurements is currently under
study by multiple groups.

Finally, although \soo\ is not always the dominate sulfur carrier in
H$_2$-dominated exoplanet atmospheres, the H$_2$S that carries most of
the sulfur is much more difficult to detect and quantify
\citep{polman:2023}.  Furthermore, since the vertical abundance
profile of photochemical \soo\ varies significantly with altitude
(Fig.~\ref{fig:metallicity}), retrievals in which the abundance is
assumed to be constant with altitude may somewhat underestimate the
true \soo\ abundance \citep[which might already affect the retrieval
  results of][]{constantinou:2023}. For all these reasons,
physically-motivated photochemical models of \soo in H$_2$-dominated
exoplanets should therefore remain a key tool to unraveling the
atmospheric properties of such planets.}



\vspace{-0.1in} \acknowledgments Dedicated to E$^3$.  We heartily
thank S.-M.\ Tsai for help with \texttt{VULCAN}, for general
discussions of photochemistry, and for useful comments on an early
draft of this paper. We thank B.\ Bitsch for several useful discussions
that improved the quality of this paper, and we thank D.\ Turrini for
clarifying several points regarding planet formation. {\update
  Finally, we also thank the anonymous referee for their useful
  comments that helped to materially improve the quality of this
  work.}


\bibliographystyle{apj}
\bibliography{../ms}

\begin{thebibliography}{}
\expandafter\ifx\csname natexlab\endcsname\relax\def\natexlab#1{#1}\fi

\bibitem[{{Adam} {et~al.}(2019){Adam}, {Yachmenev}, {Yurchenko}, \&
  {Jensen}}]{adam:2019}
{Adam}, A.~Y., {Yachmenev}, A., {Yurchenko}, S.~N., \& {Jensen}, P. 2019,
  Journal of Physical Chemistry A, 123, 4755

\bibitem[{{Ahrer} {et~al.}(2023){Ahrer}, {Stevenson}, {Mansfield}, {Moran},
  {Brande}, {Morello}, {Murray}, {Nikolov}, {Petit dit de la Roche},
  {Schlawin}, {Wheatley}, {Zieba}, {Batalha}, {Damiano}, {Goyal}, {Lendl},
  {Lothringer}, {Mukherjee}, {Ohno}, {Batalha}, {Battley}, {Bean}, {Beatty},
  {Benneke}, {Berta-Thompson}, {Carter}, {Cubillos}, {Daylan}, {Espinoza},
  {Gao}, {Gibson}, {Gill}, {Harrington}, {Hu}, {Kreidberg}, {Lewis}, {Line},
  {L{\'o}pez-Morales}, {Parmentier}, {Powell}, {Sing}, {Tsai}, {Wakeford},
  {Welbanks}, {Alam}, {Alderson}, {Allen}, {Anderson}, {Barstow}, {Bayliss},
  {Bell}, {Blecic}, {Bryant}, {Burleigh}, {Carone}, {Casewell}, {Changeat},
  {Chubb}, {Crossfield}, {Crouzet}, {Decin}, {D{\'e}sert}, {Feinstein},
  {Flagg}, {Fortney}, {Gizis}, {Heng}, {Iro}, {Kempton}, {Kendrew}, {Kirk},
  {Knutson}, {Komacek}, {Lagage}, {Leconte}, {Lustig-Yaeger}, {MacDonald},
  {Mancini}, {May}, {Mayne}, {Miguel}, {Mikal-Evans}, {Molaverdikhani},
  {Palle}, {Piaulet}, {Rackham}, {Redfield}, {Rogers}, {Roy}, {Rustamkulov},
  {Shkolnik}, {Sotzen}, {Taylor}, {Tremblin}, {Tucker}, {Turner}, {de
  Val-Borro}, {Venot}, \& {Zhang}}]{ahrer:2023}
{Ahrer}, E.-M., {Stevenson}, K.~B., {Mansfield}, M., {et~al.} 2023, \nat, 614,
  653

\bibitem[{{Alderson} {et~al.}(2023){Alderson}, {Wakeford}, {Alam}, {Batalha},
  {Lothringer}, {Adams Redai}, {Barat}, {Brande}, {Damiano}, {Daylan},
  {Espinoza}, {Flagg}, {Goyal}, {Grant}, {Hu}, {Inglis}, {Lee}, {Mikal-Evans},
  {Ramos-Rosado}, {Roy}, {Wallack}, {Batalha}, {Bean}, {Benneke},
  {Berta-Thompson}, {Carter}, {Changeat}, {Col{\'o}n}, {Crossfield},
  {D{\'e}sert}, {Foreman-Mackey}, {Gibson}, {Kreidberg}, {Line},
  {L{\'o}pez-Morales}, {Molaverdikhani}, {Moran}, {Morello}, {Moses},
  {Mukherjee}, {Schlawin}, {Sing}, {Stevenson}, {Taylor}, {Aggarwal}, {Ahrer},
  {Allen}, {Barstow}, {Bell}, {Blecic}, {Casewell}, {Chubb}, {Crouzet},
  {Cubillos}, {Decin}, {Feinstein}, {Fortney}, {Harrington}, {Heng}, {Iro},
  {Kempton}, {Kirk}, {Knutson}, {Krick}, {Leconte}, {Lendl}, {MacDonald},
  {Mancini}, {Mansfield}, {May}, {Mayne}, {Miguel}, {Nikolov}, {Ohno}, {Palle},
  {Parmentier}, {Petit dit de la Roche}, {Piaulet}, {Powell}, {Rackham},
  {Redfield}, {Rogers}, {Rustamkulov}, {Tan}, {Tremblin}, {Tsai}, {Turner}, {de
  Val-Borro}, {Venot}, {Welbanks}, {Wheatley}, \& {Zhang}}]{alderson:2023}
{Alderson}, L., {Wakeford}, H.~R., {Alam}, M.~K., {et~al.} 2023, \nat, 614, 664

\bibitem[{{Asplund} {et~al.}(2009){Asplund}, {Grevesse}, {Sauval}, \&
  {Scott}}]{asplund:2009}
{Asplund}, M., {Grevesse}, N., {Sauval}, A.~J., \& {Scott}, P. 2009, \araa, 47,
  481

\bibitem[{{Azzam} {et~al.}(2016){Azzam}, {Tennyson}, {Yurchenko}, \&
  {Naumenko}}]{azzam:2016}
{Azzam}, A. A.~A., {Tennyson}, J., {Yurchenko}, S.~N., \& {Naumenko}, O.~V.
  2016, \mnras, 460, 4063

\bibitem[{{Barber} {et~al.}(2014){Barber}, {Strange}, {Hill}, {Polyansky},
  {Mellau}, {Yurchenko}, \& {Tennyson}}]{barber:2014}
{Barber}, R.~J., {Strange}, J.~K., {Hill}, C., {et~al.} 2014, \mnras, 437, 1828

\bibitem[{{Batalha} {et~al.}(2017){Batalha}, {Bean}, {Stevenson}, {Alam},
  {Batalha}, {Benneke}, {Berta-Thompson}, {Blecic}, {Bruno}, {Carter},
  {Chapman}, {Crossfield}, {Crouzet}, {Decin}, {Demory}, {Desert}, {Dragomir},
  {Fortney}, {Fraine}, {Gao}, {Garcia Munoz}, {Gibson}, {Goyal}, {Harrington},
  {Heng}, {Hu}, {Kempton}, {Kendrew}, {Kilpatrick}, {Knutson}, {Kreidberg},
  {Krick}, {Lagage}, {Lendl}, {Line}, {Lopez-Morales}, {Louden}, {Madhusudhan},
  {Mandell}, {Mansfield}, {May}, {Mikal-Evans}, {Morello}, {Morley}, {Moses},
  {Nikolov}, {Parmentier}, {Redfield}, {Roberts}, {Schlawin}, {Showman},
  {Sing}, {Spake}, {Swain}, {Todorov}, {Tsiaras}, {Venot}, {Waalkes},
  {Wakeford}, {Wheatley}, \& {Zellem}}]{batalha:2017ers}
{Batalha}, N., {Bean}, J.~L., {Stevenson}, K.~B., {et~al.} 2017, {The
  Transiting Exoplanet Community Early Release Science Program}, JWST Proposal
  ID 1366. Cycle 0 Early Release Science

\bibitem[{{Bernath}(2020)}]{bernath:2020}
{Bernath}, P.~F. 2020, \jqsrt, 240, 106687

\bibitem[{{Biazzo} {et~al.}(2022){Biazzo}, {D'Orazi}, {Desidera}, {Turrini},
  {Benatti}, {Gratton}, {Magrini}, {Sozzetti}, {Baratella}, {Bonomo}, {Borsa},
  {Claudi}, {Covino}, {Damasso}, {Di Mauro}, {Lanza}, {Maggio}, {Malavolta},
  {Maldonado}, {Marzari}, {Micela}, {Poretti}, {Vitello}, {Affer}, {Bignamini},
  {Carleo}, {Cosentino}, {Fiorenzano}, {Giacobbe}, {Harutyunyan}, {Leto},
  {Mancini}, {Molinari}, {Molinaro}, {Nardiello}, {Nascimbeni}, {Pagano},
  {Pedani}, {Piotto}, {Rainer}, \& {Scandariato}}]{biazzo:2022}
{Biazzo}, K., {D'Orazi}, V., {Desidera}, S., {et~al.} 2022, \aap, 664, A161

\bibitem[{{Bitsch} {et~al.}(2022){Bitsch}, {Schneider}, \&
  {Kreidberg}}]{bitsch:2022}
{Bitsch}, B., {Schneider}, A.~D., \& {Kreidberg}, L. 2022, \aap, 665, A138

\bibitem[{Brooke {et~al.}(2016)Brooke, Bernath, Western, Sneden, Afşar, Li, \&
  Gordon}]{brooke:2016}
Brooke, J.~S., Bernath, P.~F., Western, C.~M., {et~al.} 2016, Journal of
  Quantitative Spectroscopy and Radiative Transfer, 168, 142

\bibitem[{{Brooke} {et~al.}(2014){Brooke}, {Ram}, {Western}, {Li}, {Schwenke},
  \& {Bernath}}]{brooke:2014}
{Brooke}, J. S.~A., {Ram}, R.~S., {Western}, C.~M., {et~al.} 2014, \apjs, 210,
  23

\bibitem[{{Chubb} {et~al.}(2020){Chubb}, {Tennyson}, \&
  {Yurchenko}}]{chubb:2020}
{Chubb}, K.~L., {Tennyson}, J., \& {Yurchenko}, S.~N. 2020, \mnras, 493, 1531

\bibitem[{{Chubb} {et~al.}(2018){Chubb}, {Naumenko}, {Keely}, {Bartolotto},
  {Macdonald}, {Mukhtar}, {Grachov}, {White}, {Coleman}, {Liu}, {Fazliev},
  {Polovtseva}, {Horneman}, {Campargue}, {Furtenbacher}, {Cs{\'a}sz{\'a}r},
  {Yurchenko}, \& {Tennyson}}]{chubb:2018}
{Chubb}, K.~L., {Naumenko}, O., {Keely}, S., {et~al.} 2018, \jqsrt, 218, 178

\bibitem[{{Constantinou} {et~al.}(2023){Constantinou}, {Madhusudhan}, \&
  {Gandhi}}]{constantinou:2023}
{Constantinou}, S., {Madhusudhan}, N., \& {Gandhi}, S. 2023, \apjl, 943, L10

\bibitem[{{Costa Silva} {et~al.}(2020){Costa Silva}, {Delgado Mena}, \&
  {Tsantaki}}]{costasilva:2020}
{Costa Silva}, A.~R., {Delgado Mena}, E., \& {Tsantaki}, M. 2020, \aap, 634,
  A136

\bibitem[{{Feinstein} {et~al.}(2023){Feinstein}, {Radica}, {Welbanks},
  {Murray}, {Ohno}, {Coulombe}, {Espinoza}, {Bean}, {Teske}, {Benneke}, {Line},
  {Rustamkulov}, {Saba}, {Tsiaras}, {Barstow}, {Fortney}, {Gao}, {Knutson},
  {MacDonald}, {Mikal-Evans}, {Rackham}, {Taylor}, {Parmentier}, {Batalha},
  {Berta-Thompson}, {Carter}, {Changeat}, {dos Santos}, {Gibson}, {Goyal},
  {Kreidberg}, {L{\'o}pez-Morales}, {Lothringer}, {Miguel}, {Molaverdikhani},
  {Moran}, {Morello}, {Mukherjee}, {Sing}, {Stevenson}, {Wakeford}, {Ahrer},
  {Alam}, {Alderson}, {Allen}, {Batalha}, {Bell}, {Blecic}, {Brande},
  {Caceres}, {Casewell}, {Chubb}, {Crossfield}, {Crouzet}, {Cubillos}, {Decin},
  {D{\'e}sert}, {Harrington}, {Heng}, {Henning}, {Iro}, {Kempton}, {Kendrew},
  {Kirk}, {Krick}, {Lagage}, {Lendl}, {Mancini}, {Mansfield}, {May}, {Mayne},
  {Nikolov}, {Palle}, {Petit dit de la Roche}, {Piaulet}, {Powell}, {Redfield},
  {Rogers}, {Roman}, {Roy}, {Nixon}, {Schlawin}, {Tan}, {Tremblin}, {Turner},
  {Venot}, {Waalkes}, {Wheatley}, \& {Zhang}}]{feinstein:2023}
{Feinstein}, A.~D., {Radica}, M., {Welbanks}, L., {et~al.} 2023, \nat, 614, 670

\bibitem[{{Fortney} {et~al.}(2013){Fortney}, {Mordasini}, {Nettelmann},
  {Kempton}, {Greene}, \& {Zahnle}}]{fortney:2013}
{Fortney}, J.~J., {Mordasini}, C., {Nettelmann}, N., {et~al.} 2013, \apj, 775,
  80

\bibitem[{{Gorman} {et~al.}(2019){Gorman}, {Yurchenko}, \&
  {Tennyson}}]{gorman:2019}
{Gorman}, M.~N., {Yurchenko}, S.~N., \& {Tennyson}, J. 2019, \mnras, 490, 1652

\bibitem[{{Harris} {et~al.}(2006){Harris}, {Tennyson}, {Kaminsky}, {Pavlenko},
  \& {Jones}}]{harris:2006}
{Harris}, G.~J., {Tennyson}, J., {Kaminsky}, B.~M., {Pavlenko}, Y.~V., \&
  {Jones}, H.~R.~A. 2006, \mnras, 367, 400

\bibitem[{{Heng} {et~al.}(2016){Heng}, {Lyons}, \& {Tsai}}]{heng:2015c}
{Heng}, K., {Lyons}, J.~R., \& {Tsai}, S.-M. 2016, \apj, 816, 96

\bibitem[{{Hobbs} {et~al.}(2021){Hobbs}, {Rimmer}, {Shorttle}, \&
  {Madhusudhan}}]{hobbs:2021}
{Hobbs}, R., {Rimmer}, P.~B., {Shorttle}, O., \& {Madhusudhan}, N. 2021,
  \mnras, 506, 3186

\bibitem[{{JTEC Team} {et~al.}(2023){JTEC Team}, {Ahrer}, {Alderson},
  {Batalha}, {Batalha}, {Bean}, {Beatty}, {Bell}, {Benneke}, {Berta-Thompson},
  {Carter}, {Crossfield}, {Espinoza}, {Feinstein}, {Fortney}, {Gibson},
  {Goyal}, {Kempton}, {Kirk}, {Kreidberg}, {L{\'o}pez-Morales}, {Line},
  {Lothringer}, {Moran}, {Mukherjee}, {Ohno}, {Parmentier}, {Piaulet},
  {Rustamkulov}, {Schlawin}, {Sing}, {Stevenson}, {Wakeford}, {Allen},
  {Birkmann}, {Brande}, {Crouzet}, {Cubillos}, {Damiano}, {D{\'e}sert}, {Gao},
  {Harrington}, {Hu}, {Kendrew}, {Knutson}, {Lagage}, {Leconte}, {Lendl},
  {MacDonald}, {May}, {Miguel}, {Molaverdikhani}, {Moses}, {Murray}, {Nehring},
  {Nikolov}, {Petit dit de la Roche}, {Radica}, {Roy}, {Stassun}, {Taylor},
  {Waalkes}, {Wachiraphan}, {Welbanks}, {Wheatley}, {Aggarwal}, {Alam},
  {Banerjee}, {Barstow}, {Blecic}, {Casewell}, {Changeat}, {Chubb},
  {Col{\'o}n}, {Coulombe}, {Daylan}, {de Val-Borro}, {Decin}, {Dos Santos},
  {Flagg}, {France}, {Fu}, {Garc{\'\i}a Mu{\~n}oz}, {Gizis}, {Glidden},
  {Grant}, {Heng}, {Henning}, {Hong}, {Inglis}, {Iro}, {Kataria}, {Komacek},
  {Krick}, {Lee}, {Lewis}, {Lillo-Box}, {Lustig-Yaeger}, {Mancini}, {Mandell},
  {Mansfield}, {Marley}, {Mikal-Evans}, {Morello}, {Nixon}, {Ortiz Ceballos},
  {Piette}, {Powell}, {Rackham}, {Ramos-Rosado}, {Rauscher}, {Redfield},
  {Rogers}, {Roman}, {Roudier}, {Scarsdale}, {Shkolnik}, {Southworth}, {Spake},
  {Steinrueck}, {Tan}, {Teske}, {Tremblin}, {Tsai}, {Tucker}, {Turner},
  {Valenti}, {Venot}, {Waldmann}, {Wallack}, {Zhang}, \& {Zieba}}]{ers:2023}
{JTEC Team}, {Ahrer}, E.-M., {Alderson}, L., {et~al.} 2023, \nat, 614, 649

\bibitem[{{Kama} {et~al.}(2019){Kama}, {Shorttle}, {Jermyn}, {Folsom},
  {Furuya}, {Bergin}, {Walsh}, \& {Keller}}]{kama:2019}
{Kama}, M., {Shorttle}, O., {Jermyn}, A.~S., {et~al.} 2019, \apj, 885, 114

\bibitem[{{Le Gal} {et~al.}(2021){Le Gal}, {{\"O}berg}, {Teague}, {Loomis},
  {Law}, {Walsh}, {Bergin}, {M{\'e}nard}, {Wilner}, {Andrews}, {Aikawa},
  {Booth}, {Cataldi}, {Bergner}, {Bosman}, {Cleeves}, {Czekala}, {Furuya},
  {Guzm{\'a}n}, {Huang}, {Ilee}, {Nomura}, {Qi}, {Schwarz}, {Tsukagoshi},
  {Yamato}, \& {Zhang}}]{legal:2021}
{Le Gal}, R., {{\"O}berg}, K.~I., {Teague}, R., {et~al.} 2021, \apjs, 257, 12

\bibitem[{{Lee} {et~al.}(2023){Lee}, {Tsai}, {Hammond}, \& {Tan}}]{lee:2023}
{Lee}, E. K.~H., {Tsai}, S.-M., {Hammond}, M., \& {Tan}, X. 2023, arXiv
  e-prints, arXiv:2302.09525

\bibitem[{{Lindzen}(1981)}]{lindzen:1981}
{Lindzen}, R.~S. 1981, \jgr, 86, 9707

\bibitem[{{Lodders}(2003)}]{lodders:2003}
{Lodders}, K. 2003, \apj, 591, 1220

\bibitem[{{Lodders}(2020)}]{lodders:2020}
---. 2020, Oxford Research Enc. of Pl. Sci., arXiv:1912.00844

\bibitem[{{Lothringer} {et~al.}(2021){Lothringer}, {Rustamkulov}, {Sing},
  {Gibson}, {Wilson}, \& {Schlaufman}}]{lothringer:2021}
{Lothringer}, J.~D., {Rustamkulov}, Z., {Sing}, D.~K., {et~al.} 2021, \apj,
  914, 12

\bibitem[{{Madhusudhan}(2012)}]{madhusudhan:2012b}
{Madhusudhan}, N. 2012, \apj, 758, 36

\bibitem[{{Masseron} {et~al.}(2014){Masseron}, {Plez}, {Van Eck}, {Colin},
  {Daoutidis}, {Godefroid}, {Coheur}, {Bernath}, {Jorissen}, \&
  {Christlieb}}]{masseron:2014}
{Masseron}, T., {Plez}, B., {Van Eck}, S., {et~al.} 2014, \aap, 571, A47

\bibitem[{{Molli{\`e}re} {et~al.}(2019){Molli{\`e}re}, {Wardenier}, {van
  Boekel}, {Henning}, {Molaverdikhani}, \& {Snellen}}]{molliere:2019}
{Molli{\`e}re}, P., {Wardenier}, J.~P., {van Boekel}, R., {et~al.} 2019, \aap,
  627, A67

\bibitem[{{Mordasini} {et~al.}(2016){Mordasini}, {van Boekel}, {Molli{\`e}re},
  {Henning}, \& {Benneke}}]{mordasini:2016}
{Mordasini}, C., {van Boekel}, R., {Molli{\`e}re}, P., {Henning}, T., \&
  {Benneke}, B. 2016, \apj, 832, 41

\bibitem[{{Moses} {et~al.}(2022){Moses}, {Tremblin}, {Venot}, \&
  {Miguel}}]{moses:2022}
{Moses}, J.~I., {Tremblin}, P., {Venot}, O., \& {Miguel}, Y. 2022, Experimental
  Astronomy, 53, 279

\bibitem[{{{\"O}berg} {et~al.}(2011){{\"O}berg}, {Murray-Clay}, \&
  {Bergin}}]{oberg:2011}
{{\"O}berg}, K.~I., {Murray-Clay}, R., \& {Bergin}, E.~A. 2011, \apjl, 743, L16

\bibitem[{{Ohno} \& {Fortney}(2022{\natexlab{a}})}]{ohno:2023a}
{Ohno}, K., \& {Fortney}, J.~J. 2022{\natexlab{a}}, arXiv e-prints,
  arXiv:2211.16876

\bibitem[{{Ohno} \& {Fortney}(2022{\natexlab{b}})}]{ohno:2023b}
---. 2022{\natexlab{b}}, arXiv e-prints, arXiv:2211.16877

\bibitem[{{Oka} {et~al.}(2011){Oka}, {Nakamoto}, \& {Ida}}]{oka:2011}
{Oka}, A., {Nakamoto}, T., \& {Ida}, S. 2011, \apj, 738, 141

\bibitem[{{Pacetti} {et~al.}(2022){Pacetti}, {Turrini}, {Schisano}, {Molinari},
  {Fonte}, {Politi}, {Hennebelle}, {Klessen}, {Testi}, \&
  {Lebreuilly}}]{pacetti:2022}
{Pacetti}, E., {Turrini}, D., {Schisano}, E., {et~al.} 2022, \apj, 937, 36

\bibitem[{{Polanski} {et~al.}(2022){Polanski}, {Crossfield}, {Howard},
  {Isaacson}, \& {Rice}}]{polanski:2022}
{Polanski}, A.~S., {Crossfield}, I. J.~M., {Howard}, A.~W., {Isaacson}, H., \&
  {Rice}, M. 2022, Research Notes of the American Astronomical Society, 6, 155

\bibitem[{{Polman} {et~al.}(2023){Polman}, {Waters}, {Min}, {Miguel}, \&
  {Khorshid}}]{polman:2023}
{Polman}, J., {Waters}, L.~B.~F.~M., {Min}, M., {Miguel}, Y., \& {Khorshid}, N.
  2023, \aap, 670, A161

\bibitem[{{Polyansky} {et~al.}(2018){Polyansky}, {Kyuberis}, {Zobov},
  {Tennyson}, {Yurchenko}, \& {Lodi}}]{polyansky:2018}
{Polyansky}, O.~L., {Kyuberis}, A.~A., {Zobov}, N.~F., {et~al.} 2018, \mnras,
  480, 2597

\bibitem[{{Rivi{\`e}re-Marichalar} {et~al.}(2022){Rivi{\`e}re-Marichalar},
  {Fuente}, {Esplugues}, {Wakelam}, {le Gal}, {Baruteau}, {Ribas},
  {Mac{\'\i}as}, {Neri}, \& {Navarro-Almaida}}]{riviere:2022}
{Rivi{\`e}re-Marichalar}, P., {Fuente}, A., {Esplugues}, G., {et~al.} 2022,
  \aap, 665, A61

\bibitem[{{Rothman} {et~al.}(2010){Rothman}, {Gordon}, {Barber}, {Dothe},
  {Gamache}, {Goldman}, {Perevalov}, {Tashkun}, \& {Tennyson}}]{rothman:2010}
{Rothman}, L., {Gordon}, I., {Barber}, R., {et~al.} 2010, JQSRT, 111, 2139

\bibitem[{{Rustamkulov} {et~al.}(2023){Rustamkulov}, {Sing}, {Mukherjee},
  {May}, {Kirk}, {Schlawin}, {Line}, {Piaulet}, {Carter}, {Batalha}, {Goyal},
  {L{\'o}pez-Morales}, {Lothringer}, {MacDonald}, {Moran}, {Stevenson},
  {Wakeford}, {Espinoza}, {Bean}, {Batalha}, {Benneke}, {Berta-Thompson},
  {Crossfield}, {Gao}, {Kreidberg}, {Powell}, {Cubillos}, {Gibson}, {Leconte},
  {Molaverdikhani}, {Nikolov}, {Parmentier}, {Roy}, {Taylor}, {Turner},
  {Wheatley}, {Aggarwal}, {Ahrer}, {Alam}, {Alderson}, {Allen}, {Banerjee},
  {Barat}, {Barrado}, {Barstow}, {Bell}, {Blecic}, {Brande}, {Casewell},
  {Changeat}, {Chubb}, {Crouzet}, {Daylan}, {Decin}, {D{\'e}sert},
  {Mikal-Evans}, {Feinstein}, {Flagg}, {Fortney}, {Harrington}, {Heng}, {Hong},
  {Hu}, {Iro}, {Kataria}, {Kempton}, {Krick}, {Lendl}, {Lillo-Box}, {Louca},
  {Lustig-Yaeger}, {Mancini}, {Mansfield}, {Mayne}, {Miguel}, {Morello},
  {Ohno}, {Palle}, {Petit dit de la Roche}, {Rackham}, {Radica},
  {Ramos-Rosado}, {Redfield}, {Rogers}, {Shkolnik}, {Southworth}, {Teske},
  {Tremblin}, {Tucker}, {Venot}, {Waalkes}, {Welbanks}, {Zhang}, \&
  {Zieba}}]{rustamkulov:2023}
{Rustamkulov}, Z., {Sing}, D.~K., {Mukherjee}, S., {et~al.} 2023, \nat, 614,
  659

\bibitem[{{Schneider} \& {Bitsch}(2021{\natexlab{a}})}]{schneider:2021a}
{Schneider}, A.~D., \& {Bitsch}, B. 2021{\natexlab{a}}, \aap, 654, A71

\bibitem[{{Schneider} \& {Bitsch}(2021{\natexlab{b}})}]{schneider:2021b}
---. 2021{\natexlab{b}}, \aap, 654, A72

\bibitem[{{Schneider} \& {Bitsch}(2022)}]{schneider:2022}
---. 2022, {How drifting and evaporating pebbles shape giant planets
  (Corrigendum)}, Astronomy \& Astrophysics, Volume 659, id.C3, 3 pp.,
  doi:10.1051/0004-6361/202141096e

\bibitem[{{Seager} {et~al.}(2005){Seager}, {Richardson}, {Hansen}, {Menou},
  {Cho}, \& {Deming}}]{seager:2005}
{Seager}, S., {Richardson}, L.~J., {Hansen}, B.~M.~S., {et~al.} 2005, \apj,
  632, 1122

\bibitem[{{Syme} \& {McKemmish}(2020)}]{syme:2020}
{Syme}, A.-M., \& {McKemmish}, L.~K. 2020, \mnras, 499, 25

\bibitem[{{Tsai} {et~al.}(2017){Tsai}, {Lyons}, {Grosheintz}, {Rimmer},
  {Kitzmann}, \& {Heng}}]{tsai:2017}
{Tsai}, S.-M., {Lyons}, J.~R., {Grosheintz}, L., {et~al.} 2017, \apjs, 228, 20

\bibitem[{{Tsai} {et~al.}(2021){Tsai}, {Malik}, {Kitzmann}, {Lyons}, {Fateev},
  {Lee}, \& {Heng}}]{tsai:2021}
{Tsai}, S.-M., {Malik}, M., {Kitzmann}, D., {et~al.} 2021, \apj, 923, 264

\bibitem[{{Tsai} {et~al.}(2023){Tsai}, {Lee}, {Powell}, {Gao}, {Zhang},
  {Moses}, {H{\'e}brard}, {Venot}, {Parmentier}, {Jordan}, {Hu}, {Alam},
  {Alderson}, {Batalha}, {Bean}, {Benneke}, {Bierson}, {Brady}, {Carone},
  {Carter}, {Chubb}, {Inglis}, {Leconte}, {Lopez-Morales}, {Miguel},
  {Molaverdikhani}, {Rustamkulov}, {Sing}, {Stevenson}, {Wakeford}, {Yang},
  {Aggarwal}, {Baeyens}, {Barat}, {Borro}, {Daylan}, {Fortney}, {France},
  {Goyal}, {Grant}, {Kirk}, {Kreidberg}, {Louca}, {Moran}, {Mukherjee},
  {Nasedkin}, {Ohno}, {Rackham}, {Redfield}, {Taylor}, {Tremblin}, {Visscher},
  {Wallack}, {Welbanks}, {Youngblood}, {Ahrer}, {Batalha}, {Behr},
  {Berta-Thompson}, {Blecic}, {Casewell}, {Crossfield}, {Crouzet}, {Cubillos},
  {Decin}, {D{\'e}sert}, {Feinstein}, {Gibson}, {Harrington}, {Heng},
  {Henning}, {Kempton}, {Krick}, {Lagage}, {Lendl}, {Line}, {Lothringer},
  {Mansfield}, {Mayne}, {Mikal-Evans}, {Palle}, {Schlawin}, {Shorttle},
  {Wheatley}, \& {Yurchenko}}]{tsai:2023}
{Tsai}, S.-M., {Lee}, E. K.~H., {Powell}, D., {et~al.} 2023, arXiv e-prints,
  arXiv:2211.10490

\bibitem[{{Turrini} {et~al.}(2021){Turrini}, {Schisano}, {Fonte}, {Molinari},
  {Politi}, {Fedele}, {Pani{\'c}}, {Kama}, {Changeat}, \&
  {Tinetti}}]{turrini:2021}
{Turrini}, D., {Schisano}, E., {Fonte}, S., {et~al.} 2021, \apj, 909, 40

\bibitem[{Underwood {et~al.}(2016)Underwood, Tennyson, Yurchenko, Huang,
  Schwenke, Lee, Clausen, \& Fateev}]{underwood:2016}
Underwood, D.~S., Tennyson, J., Yurchenko, S.~N., {et~al.} 2016, Monthly
  Notices of the Royal Astronomical Society, 459, 3890

\bibitem[{Wood {et~al.}(2019)Wood, Smythe, \& Harrison}]{wood:2019}
Wood, B.~J., Smythe, D.~J., \& Harrison, T. 2019, American Mineralogist, 104,
  844

\bibitem[{Yousefi {et~al.}(2018)Yousefi, Bernath, Hodges, \&
  Masseron}]{yousefi:2018}
Yousefi, M., Bernath, P.~F., Hodges, J., \& Masseron, T. 2018, Journal of
  Quantitative Spectroscopy and Radiative Transfer, 217, 416

\bibitem[{{Yurchenko} {et~al.}(2020){Yurchenko}, {Mellor}, {Freedman}, \&
  {Tennyson}}]{yurchenko:2020}
{Yurchenko}, S.~N., {Mellor}, T.~M., {Freedman}, R.~S., \& {Tennyson}, J. 2020,
  \mnras, 496, 5282

\bibitem[{{Yurchenko} \& {Tennyson}(2014)}]{yurchenko:2014}
{Yurchenko}, S.~N., \& {Tennyson}, J. 2014, \mnras, 440, 1649

\bibitem[{{Zahnle} {et~al.}(2009{\natexlab{a}}){Zahnle}, {Marley}, \&
  {Fortney}}]{zahnle:2009}
{Zahnle}, K., {Marley}, M.~S., \& {Fortney}, J.~J. 2009{\natexlab{a}}, ArXiv
  e-prints, arXiv:0911.0728

\bibitem[{{Zahnle} {et~al.}(2009{\natexlab{b}}){Zahnle}, {Marley}, {Freedman},
  {Lodders}, \& {Fortney}}]{zahnle:2009a}
{Zahnle}, K., {Marley}, M.~S., {Freedman}, R.~S., {Lodders}, K., \& {Fortney},
  J.~J. 2009{\natexlab{b}}, \apjl, 701, L20

\bibitem[{{Zahnle} {et~al.}(2016){Zahnle}, {Marley}, {Morley}, \&
  {Moses}}]{zahnle:2016}
{Zahnle}, K., {Marley}, M.~S., {Morley}, C.~V., \& {Moses}, J.~I. 2016, \apj,
  824, 137

\end{thebibliography}

\end{document}